\documentclass[10pt,twocolumn,letterpaper]{article}
\usepackage{cvpr}
\pdfoutput=1 
\usepackage{times}
\usepackage{epsfig}
\usepackage{graphicx}
\usepackage{booktabs}
\usepackage{multirow}
\usepackage{amsmath}
\usepackage{amssymb}
\usepackage{adjustbox}
\usepackage{xcolor}
\usepackage{mathrsfs}
\newcommand{\p}[0]{{{\beta}}}
\newcommand{\m}[0]{{{\delta}}}
\newcommand{\q}[0]{{{\mu}}}
\newcommand{\n}[0]{{{\eta}}}

\newcommand{\Diff}[0]{{\rm Diff}}

\newcommand{\opt}[0]{{\tilde{v}^{opt}}}
\newcommand{\pre}[0]{{\tilde{v}^{pre}}}

\newcommand{\Dist}[0]{{\rm Dist}}

\usepackage[breaklinks=true,bookmarks=false]{hyperref}

\cvprfinalcopy 

\def\cvprPaperID{6110} 

\begin{document}

\title{DeepFLASH: An Efficient Network for Learning-based \\ Medical Image Registration}
\author{Jian Wang\\
University of Virginia\\
{\tt\small jw4hv@virginia.edu}
\and
Miaomiao Zhang\\
University of Virginia\\
{\tt\small mz8rr@virginia.edu }
}

\maketitle

\begin{abstract}
This paper presents DeepFLASH, a novel network with efficient training and inference for learning-based medical image registration. In contrast to existing approaches that learn spatial transformations from training data in the high dimensional 
imaging space, we develop a new registration network entirely in a low dimensional bandlimited space. 
This dramatically reduces the computational cost and memory footprint of an expensive training and inference. 
To achieve this goal, we first introduce complex-valued operations and representations of neural architectures
that provide key components for learning-based registration models. We then construct an explicit loss function of transformation fields fully characterized in a bandlimited space with much fewer parameterizations. Experimental results show that our method is
significantly faster than the state-of-the-art deep learning based image registration methods, while producing equally accurate alignment. 
We demonstrate our algorithm in two different applications of image registration: 2D synthetic data and 3D real brain magnetic resonance (MR) images. Our code is available at \url{https://github.com/jw4hv/deepflash}.
\end{abstract}

\section{Introduction}
Image registration has been widely used in medical image analysis, for instance, atlas-based image segmentation~\cite{ashburner2005unified,Joshi2004}, anatomical shape analysis~\cite{niethammer2011geodesic,zhang2016low}, and motion correction in dynamic imaging~\cite{liao2016temporal}. The problem of deformable image registration is typically formulated as an optimization, seeking for a nonlinear and dense (voxelwise) spatial transformation between images. In many applications, it is desirable that such transformations be diffeomorphisms, i.e., differentiable, bijective mappings with differentiable inverses. In this paper, we focus on diffeomorphic image registration highlighting with a set of critical features: (i) it captures large deformations that often occur in brain shape variations, lung motions, or fetal movements; (ii) the topology of objects in the image remain intact; and (iii) no non-differentiable artifacts, such as creases or sharp corners, are introduced. However, achieving an optimal solution of diffeomorphic image registration, especially for large-scale or high-resolution images (e.g., a 3D brain MRI scan with the size of $256^3$), is computationally intensive and time-consuming. 

Attempts at speeding up diffeomorphic image registration have been made in recent works by improving numerical approximation schemes. For example, Ashburner and Friston~\cite{ashburner2011diffeomorphic} employ a Gauss-Newton method to accelerate the convergence of large deformation diffeomorphic metric mapping (LDDMM) algorithm~\cite{beg2005computing}. Zhang et al. propose a low dimensional approximation of diffeomorphic transformations, resulting in fast computation of the gradients for iterative optimization~\cite{zhang2017frequency,wang2019data}. While these methods have led to substantial reductions in running time, such gradient-based optimization still takes minutes to finish. Instead of minimizing a complex registration energy function~\cite{beg2005computing,zhang2019fast}, an alternative approach has leveraged deep learning techniques to improve registration speed by building prediction models of transformation parameters. Such algorithms typically adopt convolutional neural networks (CNNs) to learn a mapping between pairwise images and associated spatial transformations in training dataset~\cite{yang2017quicksilver,rohe2017svf,balakrishnan2019voxelmorph,chou20132d,cao2015semi}. Registration of new testing images is then achieved rapidly by evaluating the learned mapping on given volumes. While the aforementioned deep learning approaches are able to fast predict the deformation parameters in testing, the training process is extremely slow and memory intensive due to the high dimensionality of deformation parameters in imaging space. In addition, enforcing the smoothness constraints of transformations when large deformation occurs is challenging in neural networks. 

To address this issue, we propose a novel learning-based registration framework DeepFLASH in a low dimensional bandlimited space, where the diffeomorphic transformations are fully characterized with much fewer dimensions. 
Our work is inspired by a recent registration algorithm FLASH (Fourier-approximated Lie Algebras for Shooting)~\cite{zhang2019fast} with the novelty of (i) developing a learning-based predictive model that further speeds up the current registration algorithms; (ii) defining a set of complex-valued operations (e.g., complex convolution, complex-valued rectifier, etc.) and complex-valued loss function of transformations entirely in a bandlimited space; and (iii) proving that our model can be easily implemented by a dual-network in the space of real-valued functions with a careful design of the network architecture. To the best of our knowledge, we are the first to introduce the low dimensional Fourier representations of diffeomorphic transformations to learning-based registration algorithms. In contrast to traditional methods that learn spatial transformations in a high dimensional imaging space, our method dramatically reduces the computational complexity of the training process where iterative computation of gradient terms are required. This greatly alleviates the problem of  time-consuming and expensive training for deep learning based registration networks. Another major benefit of DeepFLASH is that the smoothness of diffeomorphic transformations is naturally preserved in the bandlimited space with low frequency components. Note that while we implement DeepFLASH in the context of convolutional neural network (CNN), it can be easily adapted to a variety of other neural networks, such as fully connected network (FCN), or recurrent neural network (RNN). We demonstrate the effectiveness of our model in both 2D synthetic and 3D real brain MRI data. 

\section{Background: Deformable Image Registration}
In this section, we briefly review the fundamentals of deformable image registration in the setting of LDDMM algorithm with geodesic shooting~\cite{vialard2012,younes2009evolutions}.
While this paper focuses on LDDMM, the theoretical tool developed in our model is widely applicable to various registration frameworks, for example, stationary velocity fields.

Consider a source image $S$ and a target image $T$ as square-integrable functions defined on a torus domain $\Omega = \mathbb{R}^d / \mathbb{Z}^d$ ($S(x), T(x) : \Omega \rightarrow \mathbb{R}$). The problem of diffeomorphic image registration is to find the shortest path (also known as geodesic) of diffeomorphic transformations
$\phi_t \in \Diff(\Omega): \Omega \rightarrow \Omega, t \in [0, 1]$, such that the deformed image $S \circ \psi_1$ at time point $t=1$ is similar to $T$. An explicit energy function of LDDMM with geodesic shooting is formulated as an image matching term plus a regularization term that guarantees the smoothness
of the transformation fields
\begin{align}
\label{eq:lddmm}
E(v_0) &= \frac{\gamma}{2} \, \Dist (S \circ  \psi_1, T) + \frac{1}{2}(L v_0, v_0), \, s.t., \nonumber \\
\frac{d\psi_t}{dt} &= - \mathcal{D} \psi_t \cdot v_t,
\end{align} 
 where $\gamma$ is a positive weight parameter, $\mathcal{D}$ denotes a Jacobian matrix, and $\Dist(\cdot, \cdot)$ is a distance function that measures the similarity between images. Commonly used distance metrics include sum-of-squared difference (L2-norm) of image intensities~\cite{beg2005computing}, normalized cross correlation (NCC)~\cite{avants2008symmetric}, and mutual information (MI)~\cite{leventon1997multiple}. The deformation $\phi$ is defined as an integral flow of the time-varying Eulerian velocity field $v_t$ that lies
in the tangent space of diffeomorphisms $V = T \Diff(\Omega)$. Here $L : V \rightarrow V^*$ is a symmetric, positive-definite differential operator that maps a tangent vector $v \in V$ into the dual space $m \in V^*$, with its inverse $K : V^* \rightarrow V$. The $(\cdot, \cdot)$ denotes a paring of a momentum vector $m \in V^*$ with a tangent vector $v \in V$.   
 
\begin{figure}[!h]
\begin{center}
 \includegraphics[width=.5\textwidth]{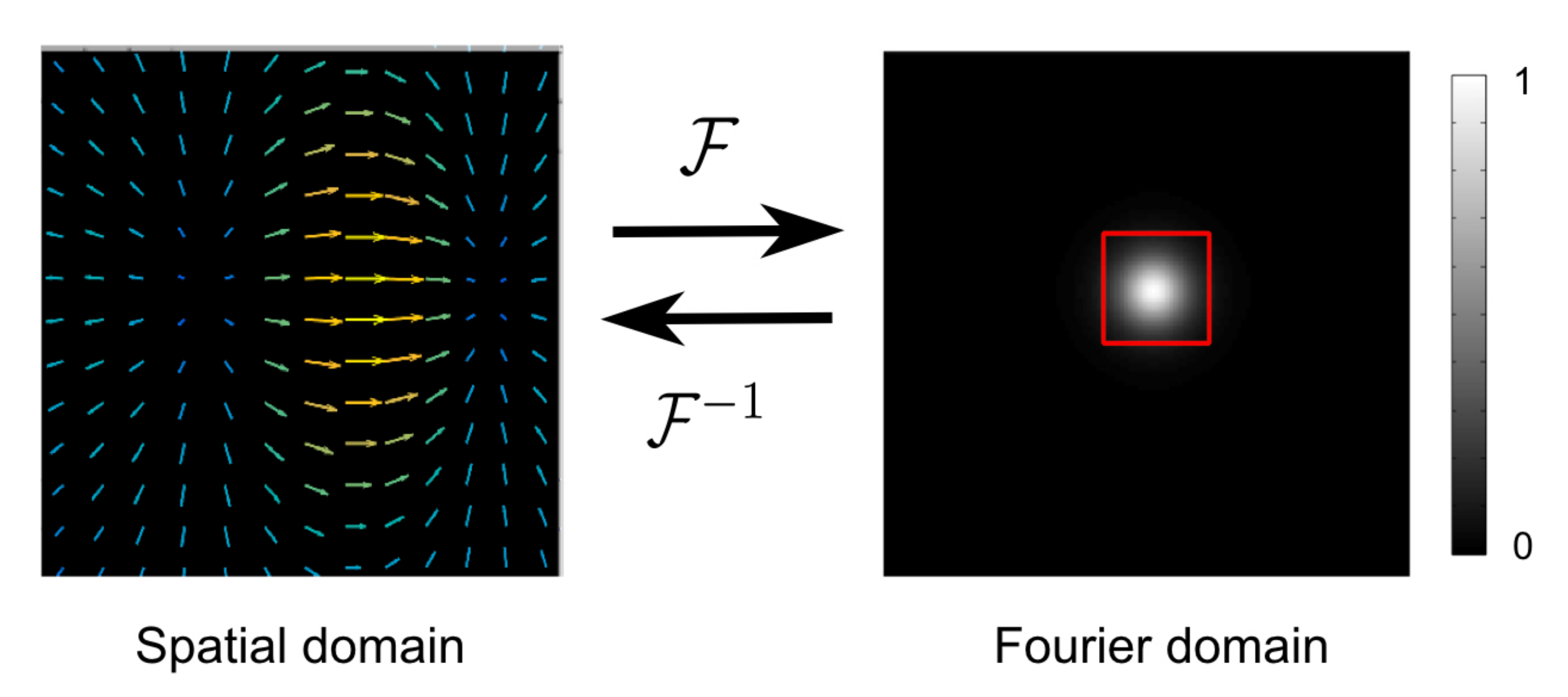}
   \caption{An example of velocity field in spatial domain vs. Fourier domain.}
\label{fig:frequency}
\end{center}
\end{figure}

The geodesic at the minimum of~\eqref{eq:lddmm} is uniquely determined by integrating the geodesic constraint, a.k.a. Euler-Poincar\'{e} differential equation (EPDiff)~\cite{arnold1966,miller2006}, 
which is computationally expensive in high dimensional image spaces. 
A recent model FLASH demonstrated that the entire optimization of LDDMM
with geodesic shooting can be efficiently carried in a low dimensional bandlimited space with dramatic speed improvement~\cite{zhang2019fast}. This is based on the fact that the velocity fields do not develop high frequencies in the Fourier domain (as shown in Fig.~\ref{fig:frequency}). We briefly review the basic concepts below. 

Let $\widetilde{\Diff}(\Omega)$ and $\tilde{V}$ denote the space of Fourier representations of diffeomorphisms
and velocity fields respectively. Given time-dependent velocity field $\tilde{v}_t \in \tilde{V}$, the diffeomorphism $\tilde{\psi}_t \in \widetilde{\Diff}(\Omega)$ in the finite-dimensional Fourier domain can be computed as
\begin{align}\label{eq:finalleftinvariantfft}
\tilde{\psi}_t = \tilde{e} + \tilde{u}_t, \quad \frac{d \tilde{u}_t}{dt} &= -\tilde{v}_t - \tilde{\mathcal{D}} \tilde{u}_t \ast \tilde{v}_t,
\end{align}
where $\tilde{e}$ is the frequency of an identity element, $\tilde{\mathcal{D}}\tilde{u}_t$ is a tensor product $\tilde{\mathcal{D}} \otimes \tilde{u}_t$, representing the Fourier frequencies of a Jacobian matrix $\tilde{\mathcal{D}}$ with central difference approximation, and $\ast$ is a circular convolution with zero padding to avoid aliasing~\footnote{To prevent the domain from growing infinity, we truncate the output of the convolution in each dimension to a suitable finite set.}.

The Fourier representation of the geodesic constraint (EPDiff) is defined as
\begin{align}\label{eq:epdiffleft}
    \frac{\partial \tilde{v}_t}{\partial t} =-\tilde{K}\left[(\tilde{\mathcal{D}} \tilde{v}_t)^T \star \tilde{m}_t + \tilde{\nabla} \cdot (\tilde{m}_t \otimes \tilde{v}_t) \right],
\end{align}
where $\star$ is the truncated matrix-vector field auto-correlation. The operator $\tilde{\nabla} \cdot$ is the discrete divergence of a vector field. Here $\tilde{K}$ is a smoothing operator with its inverse $\tilde{L}$, which is the Fourier transform of a commonly used Laplacian operator $(-\alpha \Delta + I)^c$, with a positive weight parameter $\alpha$ and a smoothness parameter~$c$. The Fourier coefficients of $\tilde{L}$ is, i.e.,
$\tilde{{L}}(\xi_1 , \ldots, \xi_d) = \left(-2 \alpha \sum_{j = 1}^d \left(\cos (2\pi \xi_j) - 1 \right) + 1\right)^c$, where $(\xi_1 , \ldots, \xi_d)$ is a d-dimensional frequency vector.

\section{Our Method: DeepFLASH}
We introduce a learning-based registration network DeepFLASH in a low dimensional bandlimited space $\tilde{V}$, with newly defined operators and functions in a complex vector space $\mathbb{C}^n$. Since the spatial transformations can be uniquely determined by an initial velocity $\tilde{v}_0$ (as introduced in Eq.~\eqref{eq:epdiffleft}), we naturally integrate this new parameterization in the architecture of DeepFLASH. To simplify the notation, we drop the time index of $\tilde{v}_0$ in remaining sections. 

Analogous to~\cite{yang2017quicksilver}, we use optimal registration results, denoted as $\tilde{v}^{opt}$, estimated by numerical optimization of the LDDMM algorithm as part of the training data. Our goal is then to predict an initial velocity $\tilde{v}^{pre}$ from image patches of the moving and target images. Before introducing DeepFLASH, we first define a set of complex-valued operations and functions that provide key components of the neural architecture. 

Consider a $Q$-dimensional complex-valued vector of input signal $\tilde{X}$ and a real-valued kernel $H$, we have a complex-valued convolution as 
\begin{align}
    H \ast \tilde{X} = H \ast \mathscr{R}(\tilde{X}) + i H \ast \mathscr{I}(\tilde{X}),
    \label{Eq:complexconv}
\end{align}
where $\mathscr{R} (\cdot)$ denotes the real part of a complex-valued vector, and $\mathscr{I} (\cdot)$ denotes an imaginary part.  

Following a recent work on complex networks~\cite{trabelsi2017deep}, we define a complex-valued activation function based on Rectified Linear Unit (ReLU). We apply real-valued ReLU separately on both of the real and imaginary part of a neuron $\tilde{Y}$ in the output layer, i.e.,
\begin{align}
 \mathbb{C}\text{ReLU}(\tilde{Y}) &= \text{ReLU}(\mathscr{R}(\tilde{Y})) + i \text{ReLU}(\mathscr{I}(\tilde{Y})).
    \label{Eq:activ} 
\end{align}

\subsection{Loss function}
Let a labeled training dataset including pairwise images and their associated optimal initial velocity fields be $\{ S_n, T_n, \tilde{v}^{opt}_n\}^{N}_{n = 1}$, where $N$ is the number of pairwise images. We model a prediction function $f(S_n, T_n; W)$ by using a convolutional neural network (CNN), with $W$ being a real-valued weight matrix for convolutional layers. We then define a loss function $\ell$ as
\begin{equation}
\ell(W) = \sum_{n=0}^{N} \| \opt_n - f(S_n, T_n; W) \| ^2_{L_2}  + \lambda \cdot  \text{Reg}(W),
\label{eq:OriginalLoss}
\end{equation}
where $\lambda$ is a positive parameter balancing between function $f$ and a regularity term $\text{Reg}(\cdot)$ on the weight matrix $W$. While we use $L_2$ norm as regularity in this paper, it is generic to other commonly used operators such as $L_1$, or $L_0$ norm.

The optimization problem of Eq.~\eqref{eq:OriginalLoss} is typically solved by using gradient-based methods. The weight matrix $W$ is updated by moving in the direction of the loss function’s steepest descent, found by its gradient $\nabla_W \ell$.

\subsection{Network Architecture: Decoupled Complex-valued Network}
\label{Sec:arch}
While we are ready to design a complex-valued registration network based on CNN, the implementation of such network is not straightforward. Trabelsi et al. developed deep complex networks to specifically handle complex-valued inputs at the cost of computational efficiency~\cite{trabelsi2017deep}. In this section, we present an efficient way to decouple the proposed complex-valued registration network into a combination of regular real-valued networks. More specifically, we construct a dual-net that separates the real and imaginary part of complex-valued network in an equivalent manner.  

Given the fact that the computation of real and imaginary parts in the convolution (Eq.~\eqref{Eq:complexconv}) and the activation function (Eq.~\eqref{Eq:activ}) are separable, we next show that the loss defined in Eq.~\eqref{eq:OriginalLoss}
can be equivalently constructed by the real and imaginary part individually. 
To simplify the notation, we define the predicted initial velocity $\pre_n \overset{\Delta}{=} f(S_n, T_n; W)$ and rewrite Eq.~\eqref{eq:OriginalLoss} as
\begin{equation}
\ell(W) = \sum_{n=0}^{N} \| \opt_n - \pre_n \| ^2_{L_2}  + \lambda \cdot  \text{Reg}(W).
\label{eq:Loss}
\end{equation}

Let $\opt_n = \p_n + i\q_n$ and $\pre_n = \m_n + i \n_n$ in Eq.~\eqref{eq:Loss}, we then obtain
\begin{align}
 \ell(W) &= \sum_{n=0}^{N} \| (\p_n + i \q_n ) - (\m_n + i \n_n)\|^2_{L_2} + \lambda \cdot  \text{Reg}(W), \notag \\
 &= 
  \sum_{n=0}^{N} \| (\p_n -\m_n ) + i (\q_n -\n_n ) \|^2_{L_2} + \lambda \cdot \text{Reg}(W), \notag \\
 &= 
  \sum_{n=0}^{N} 
  \| \p_n -\m_n \|_{L_2} ^2 + \| \q_n -\m_n \|_{L_2}^2 + \lambda \cdot \text{Reg}(W), \notag \\
 &= 
  \sum_{n=0}^{N} \| \mathscr{R}(\opt_n) - \mathscr{R}(\pre_n) \| ^2_{L_2} \notag \\ \quad &+  
  \| \mathscr{I}(\opt_n) - \mathscr{I}(\pre_n) \| ^2_{L_2}  + 
  \lambda \cdot  \text{Reg}(W).
 \label{Eq:OrginalLossCom}
 \end{align}
 
\begin{figure*}[h!]
\label{netarch}
\begin{center}
 \includegraphics[width=1.0\textwidth] {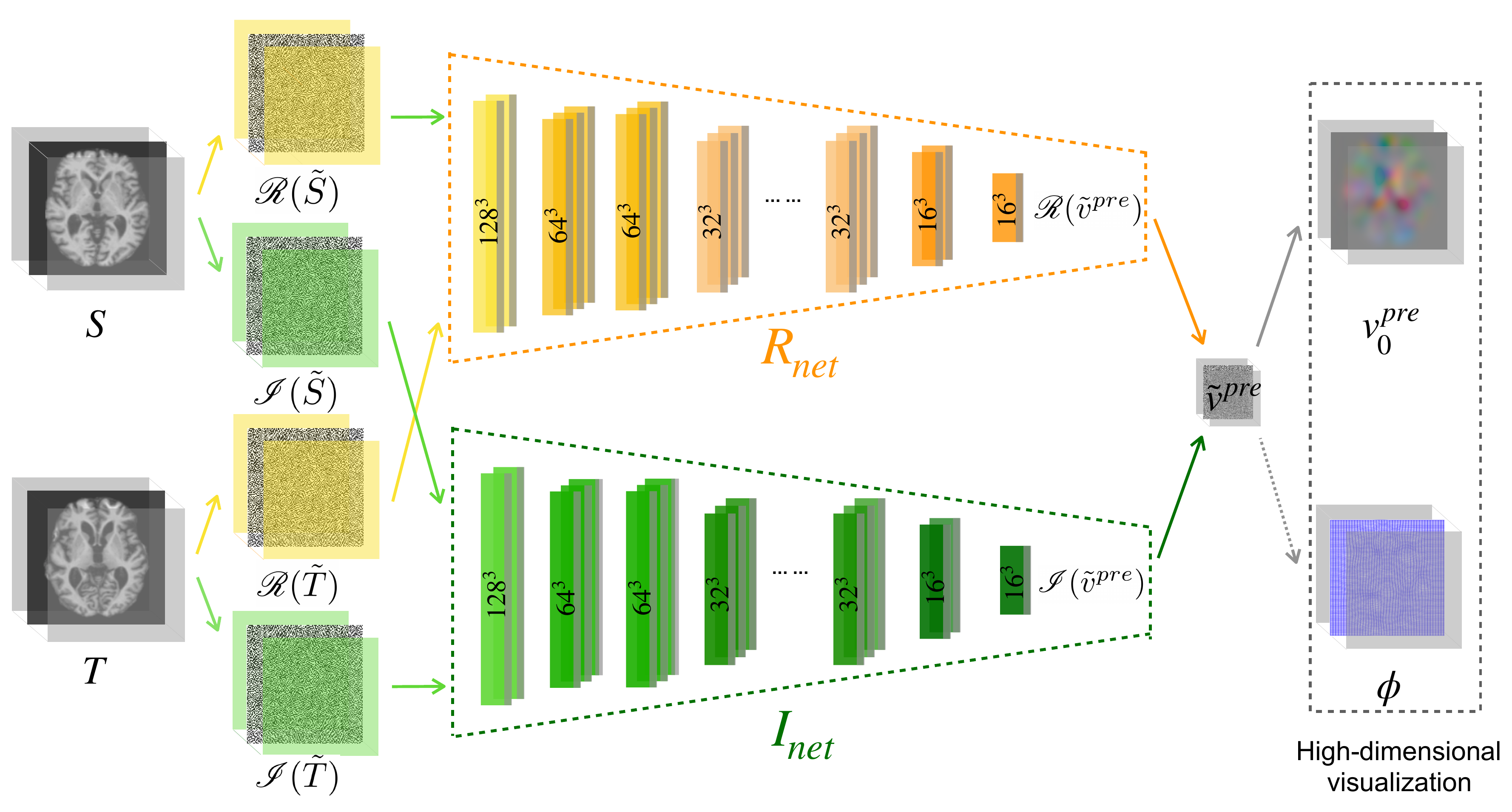}
     \caption{Architecture of DeepFLASH with dual net: S and T denote the source and target image from high dimensional spatial domain. $\mathscr{R}(\tilde{S})$ and $\mathscr{I}(\tilde{S})$ are the real and imaginary frequency spectrum convert from S. $\mathscr{R}(\tilde{T})$ and $\mathscr{I}(\tilde{T})$ are the real and imaginary frequency convert from T. $\pre$ is the low dimensional prediction optimized from our model. $v_0^{pre}$ and $\phi$ are the velocity and transformation field we recovered from our low dimensional prediction. }
\label{fig:netarch}
\end{center}
\end{figure*}

Fig.~\ref{fig:netarch} visualizes the architecture of our proposed model DeepFLASH. The input source and target images are defined in the Fourier space with the real part of frequencies as $\mathscr{R}(\tilde{S})$ and $\mathscr{R}(\tilde{T})$ vs. the imaginary parts as $\mathscr{I}(\tilde{S})$ and $\mathscr{I}(\tilde{T})$. We train real and imaginary parts in two individual neural networks $\text{R}_{net}$ and $\text{I}_{net}$. The optimization stays in a low dimensional bandlimited space without converting the coefficients ($\mathscr{R}(\pre)$, $\mathscr{I}(\pre)$) to high dimensional imaging domain back and forth. It is worthy to mention that our decoupled network is not constrained by any specific architecture. The CNN network in Fig.~\ref{fig:netarch} can be easily replaced by a variety of the state-of-the-art models, e.g., U-net~\cite{ronneberger2015u}, or fully connected neural network.

\subsection{Computational complexity}
\label{Sec:complexity}
It has been previously shown that the time complexity of convolutional layers~\cite{he2015convolutional} is $O(\sum^{P}_{p=1}b_{p-1} \cdot h^2_p \cdot b_p \cdot {Z}^2_p)$, where $p$ is the index of a convolutional layer and $P$ denotes the number of convolutional layers. The $b_{p-1}$, $b_{p}$ $h_p$ are defined as the number of input channels, output channels and the kernel size of the $p$-th layer respectively. Here $Z_p$ is the output dimension of the $p$-th layer, i.e., $Z_p = 128^3$ when the last layer predicts transformation fields for 3D images with the dimension of $128^3$. 

Current learning approaches for image registration have been performed in the original high dimensional image space~\cite{balakrishnan2019voxelmorph,yang2017quicksilver}. In contrast, our proposed model DeepFLASH significantly reduces the dimension of $Z_p$ into a low dimensional bandlimited space $z_p$ (where $z_p \ll Z_p$). This makes the training of traditional registration networks much more efficient in terms of both time and memory consumption. 

\section{Experimental Evaluation}
We demonstrate the effectiveness of our model by training and testing on both 2D synthetic data and 3D real brain MRI scans.

\subsection{Data}
\paragraph{2D synthetic data.} We first simulate $3000$ ``bull-eye" synthetic data (as shown in Fig~\ref{fig:synthetic}) by manipulating the width $a$ and height $b$ of an ellipse equation, formulated as $\frac{(x-50)^2}{a^2} + \frac{(y-50)^2}{b^2} = 1$. We draw the parameter $a, b$ randomly from a Gaussian distribution $\mathcal{N}(4,2^{2})$ for inner ellipse, and $a, b \sim \mathcal{N}(13,4^{2})$ for outer ellipse. 

\paragraph{3D brain MRI.} We include $3200$ public T1-weighted 3D brain MRI scans from Alzheimer’s Disease Neuroimaging Initiative (ADNI)~\cite{jack2008alzheimer}, Open Access Series of Imaging Studies (OASIS)~\cite{fotenos2005normative}, Autism Brain Imaging Data Exchange (ABIDE)~\cite{di2014autism}, and LONI Probabilistic Brain Atlas Individual Subject Data (LPBA40)~\cite{shattuck2008construction} with $1000$ subjects. Due to the difficulty of preserving the diffeomorphic property across individual subjects particularly with large age variations, we carefully evaluate images from subjects aged from 60 to 90. All MRIs were all pre-processed as $128\times128\times128$, $1.25mm^{3}$ isotropic voxels, and underwent skull-stripped, intensity normalized, bias field corrected and pre-aligned with affine transformation. 

To further validate our model accuracy through segmentation labels, we use manually delineated anatomical structures in LPBA40 dataset. We randomly select $100$ pairs of MR images with segmentation labels from LPBA40. We then carefully down-sampled all images and labels from the dimension of $181\times217\times181$ to $128\times128\times128$. 

\subsection{Experiments}
To validate the registration performance of DeepFLASH on both 2D and 3D data, we run a recent state-of-the-art image registration algorithm FLASH~\cite{zhang2015fast,zhang2017frequency} on $1000$ 2D synthetic data and $2000$ pairs of 3D MR images to optimal solutions, which are used as ground truth. We then randomly choose 500 pairs of 2D data and 1000 pairs of 3D MRIs from the rest of the data as testing cases. We set registration parameter $\alpha=3, c=6$ for the operator $\tilde{\ell}$, the number of time steps for Euler integration in geodesic shooting as $10$. We adopt the band-limited dimension of the initial velocity field $\opt$ as $16$, which has been shown to produce comparable registration accuracy~\cite{zhang2015fast}. We set the batch size as $64$ with learning rate $\eta = 1e-4$ for network training, then run $2000$ epochs for 2D synthetic training and $5000$ epochs for 3D brain training.

Next, we compare our 2D prediction with registration performed in full-spatial domain. For 3D testing, we compare our method with three baseline algorithms, including FLASH~\cite{zhang2019fast} (a fast optimization-based image registration method), Voxelmorph~\cite{balakrishnan2018unsupervised} (an unsupervised registration in image space) and Quicksilver~\cite{yang2017quicksilver} (a supervised method that predicts transformations in a momentum space). We also compare the registration time of DeepFLASH with traditional optimization-based methods, such as vector momenta LDDMM (VM-LDDMM)~\cite{singh2013vector}, and symmetric diffeomorphic image registration with cross-correlation (SyN) from ANTs~\cite{avants2011reproducible}. For fair comparison, we train all baseline algorithms on the same dataset and report their best performance from published source codes (e.g., \url{https://github.com/rkwitt/quicksilver}, \url{https://github.com/balakg/voxelmorph}). 

To better validate the transformations generated by DeepFLASH, we perform registration-based segmentation and examine the resulting segmentation accuracy over eight brain structures, including Putamen (Puta), Cerebellum (Cer), Caudate (Caud), Hippocampus (Hipp), Insular cortex (Cor), Cuneus (Cune), brain stem (Stem) and superior frontal gyrus (Gyrus). We evaluate a volume-overlapping similarity measurement, also known as S$\o$rensen$-$Dice coefficient, between the propagated segmentation and the manual segmentation~\cite{dice1945measures}.

We demonstrate the efficiency of our model by comparing quantitative time and GPU memory consumption across all methods. All optimal solutions for training data are generated on an i7, 9700K CPU with 32 GB internal memory. The training and prediction procedure of all learning-based methods are performed on Nvidia GTX 1080Ti GPUs. 

\subsection{Results}
\begin{figure*}[h]
\centering
 \includegraphics[width=1.0\textwidth] {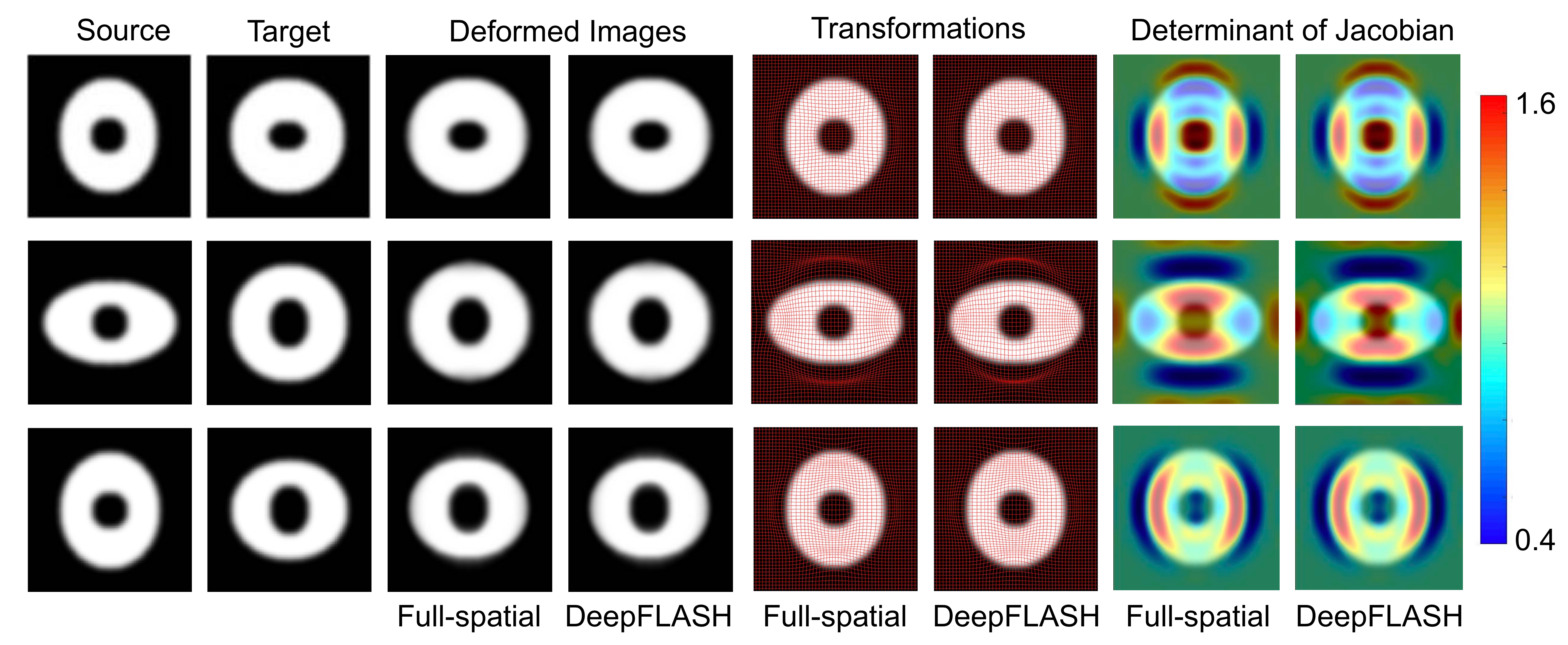}
     \caption{Example of 2D registration results. Left to right: 2D synthetic source, target image, deformed image computed in the full-spatial domain, transformation grids overlaid with source image, and determinant of Jacobian (DetJac) of transformations.}
\label{fig:synthetic}
\end{figure*}

Fig.\ref{fig:synthetic} visualizes the deformed images, transformation fields, and the determinant of Jacobian (DetJac) of transformations estimated by DeepFLASH and a registration method that performed in full-spatial image domain. Note that the value of DetJac indicates how volume changes, for instance, there is no volume change when $\text{DetJac}=1$, while volume shrinks when $\text{DetJac}<1$ and expands when $\text{DetJac}>1$. The value of DetJac smaller than zero indicates an artifact or singularity in the transformation field, i.e., a failure to preserve the diffeomorphic property when the effect of folding occurs. Both methods show similar patterns of volume changes over transformation fields. Our predicted results are fairly close to the estimates from registration algorithms in full spatial domain. 

\begin{figure*}[!h]
\begin{center}
 \includegraphics[width=1.0\textwidth] {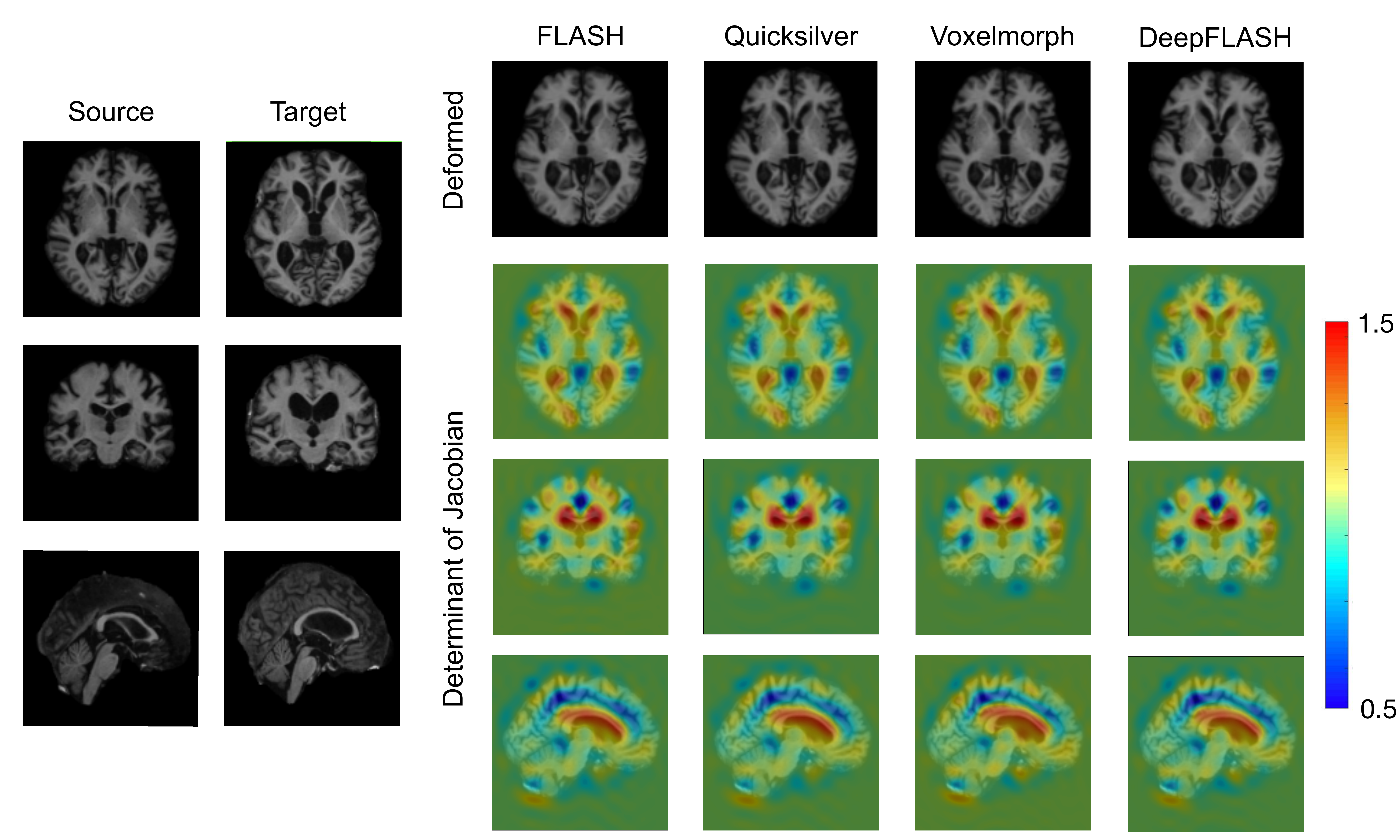}
     \caption{Example of 3D image registration on OASIS dataset. Left panel: axial, coronal, and sagittal view of source and target images. Right panel: deformed images and determinant of Jacobian of the transformations by FLASH, Quicksilver, Voxelmorph, and DeepFLASH. }
\label{fig:3Dbrain}
\end{center}
\end{figure*}

Fig.\ref{fig:3Dbrain} visualizes the deformed images and the determinant of Jacobian with pairwise registration on 3D brain MRIs for all methods. It demonstrates that our method DeepFLASH is able to produce comparable registration results with little to no loss of the accuracy. In addition, our method gracefully guarantees the smoothness of transformation fields without artifacts and singularities.   

\begin{figure*}[!htb]
\begin{center}
    \begin{minipage}{0.5\linewidth}
    \begin{flushright}
        \includegraphics[width=0.85\textwidth] {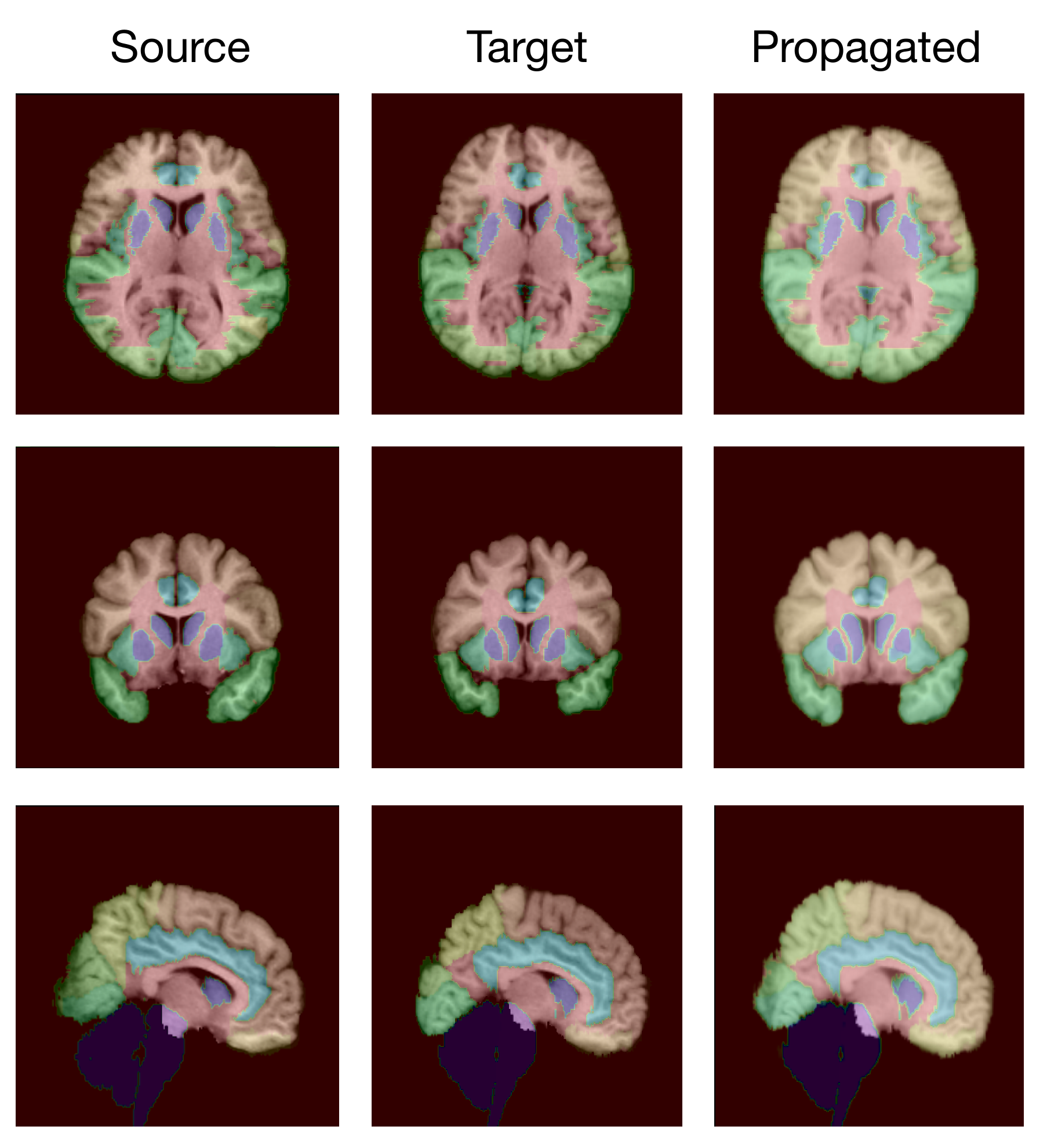}
    \end{flushright}    
    \end{minipage}
    \hspace{-1.8 cm}
    \begin{minipage}{0.49\linewidth}
    \centering
    \begin{tabular}{cc}
\hline
\multirow{2}{*}{Methods} & \multirow{2}{*}{Average Dice} \\
                         &                               \\ \hline
VM-LDDMM                 & 0.760                         \\
ANTs(SyN)                & 0.770                         \\
FLASH                    & 0.788                         \\ \hline
Quicksilver              & 0.762                         \\
Voxelmorph               & 0.774                         \\
DeepFLASH                & 0.780                         \\ \hline
\end{tabular}
    \end{minipage}
    \vspace{3mm}
\caption{Left: source and target segmentations with manually annotated eight anatomical structures (on LPBA40 dataset), propagated segmentation label deformed by our method. Right: quantitative result of average dice for all methods.}
\label{fig:seglabel}
\end{center}
\end{figure*}

The left panel of Fig.\ref{fig:seglabel} displays an example of the comparison between the manually labeled segmentation and propagated segmentation deformed by transformation field estimated from our algorithm. It clearly shows that our generated segmentations align fairly well with the manual delineations. The right panel of Fig.\ref{fig:seglabel} compares quantitative results of the average dice for all methods, with the observations that our method slightly outperforms the baseline algorithms.

\begin{figure*}[!htb]
\begin{center}
 \includegraphics[width=1\textwidth] {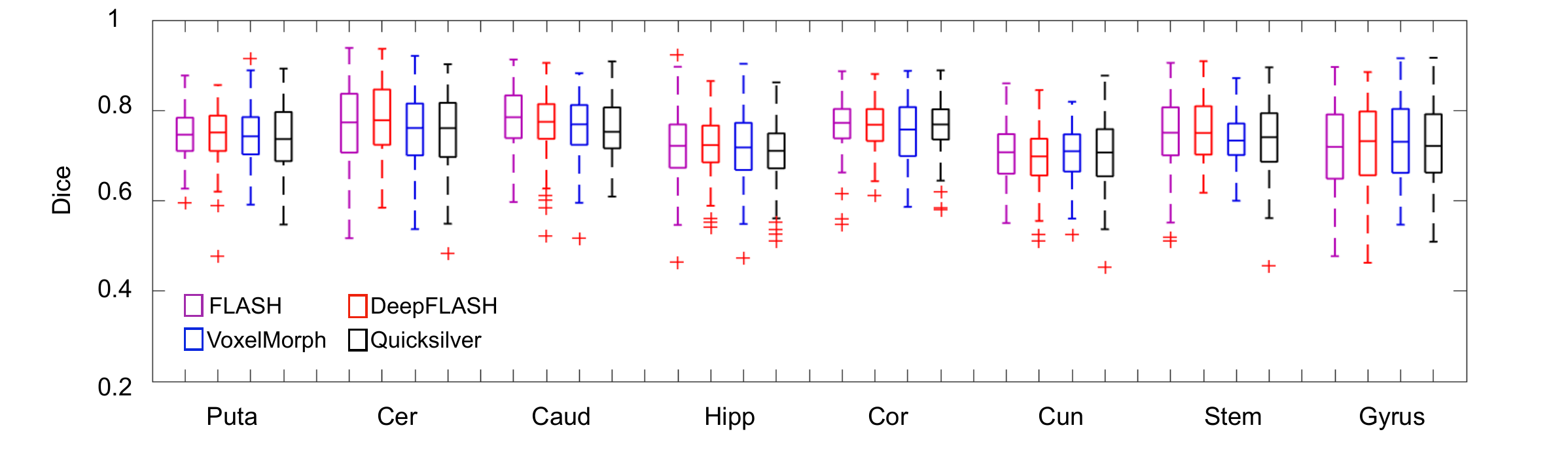}
     \caption{ Dice score evaluation by propagating the deformation field to the segmentation labels for four methods on eight brain structures (Putamen(Puta), Cerebellum (Cer), Caudate (Caud), Hippocampus (Hipp), Insular cortex (Cor), Cuneus (Cune), brain stem (Stem), and superior frontal gyrus (Gyrus)).}
\label{fig:dice}
\end{center}
\end{figure*}

Fig.~\ref{fig:dice} reports the statistics of dice scores (mean and variance) of all methods over eight brain structures from $100$ registration pairs. Our method DeepFLASH produces comparable dice scores with significantly fast training of the neural networks.

\begin{figure}[!htb]
\begin{center}
 \includegraphics[width=.48\textwidth] {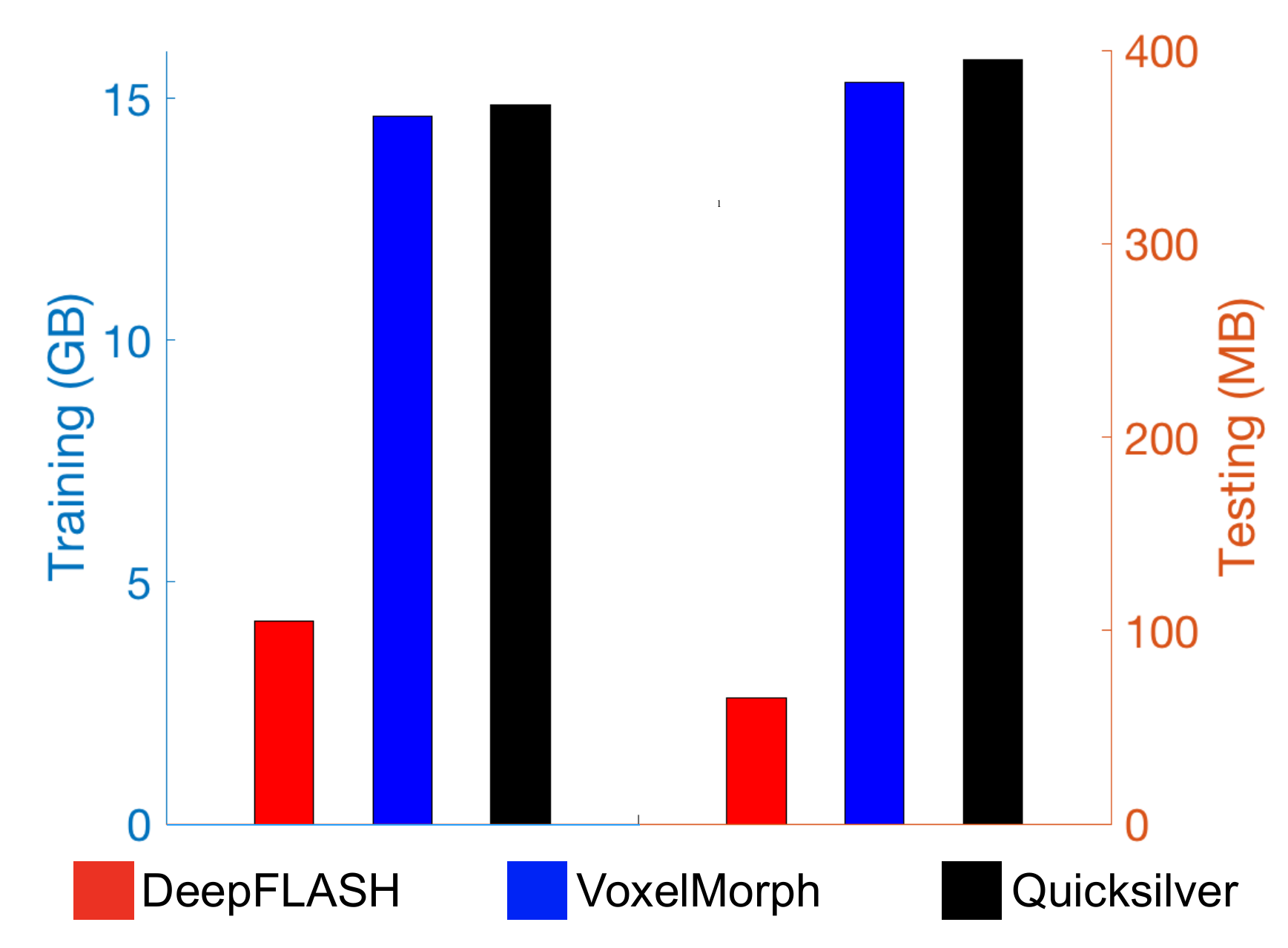}
     \caption{Comparison of average GPU memory usage on our method and learning-based baselines for both training and testing processes.}
\label{fig:memory}
\end{center}
\end{figure}

Table.\ref{Tab:Time} reports the time consumption of optimization-based registration methods, as well as the training and testing time of learning-based registration algorithms. Our method can predict the transformation fields between images approximately $100$ times faster than optimization-based registration methods. In addition, DeepFLASH outperforms learning-based registration approaches in testing, while with significantly reduced computational time in training. 

\begin{table}[!htb]
\small
\caption{Top: quantitative results of time consumption on both CPU and GPU for optimization-based registration methods (there is no GPU-version of ANTs). Bottom: computational time of training and prediction for our method DeepFLASH and current deep learning registration approaches.}
\begin{center}
\begin{tabular}{cccc}
\hline
\multirow{2}{*}{Methods} & \multirow{2}{*}{\begin{tabular}[c]{@{}c@{}}Training Time \\ (GPU hours)\end{tabular}} & \multicolumn{2}{c}{\begin{tabular}[c]{@{}c@{}}Registration time\\ (sec)\end{tabular}} \\ \cline{3-4} & & CPU & GPU \\ \hline
VM-LDDMM & - & 1210 & 262  \\
ANTs(SyN) & - & 6840 & -  \\
FLASH      & -& 286          & 53.4  \\ \hline
Quicksilver & 31.4 & 122      & 0.760 \\
Voxelmorph  & 29.7            & 52   & 0.571 \\
DeepFLASH   & \textbf{14.1}   & \textbf{41}  & \textbf{0.273}  \\ \hline
\end{tabular}
\label{Tab:Time}
\end{center}
\end{table}

Fig.\ref{fig:memory} provides the average training and testing GPU memory usage across all methods. It has been shown that our proposed method dramatically lowers the GPU footprint compared to other learning-based methods in both training and testing. 

\section{Conclusion}
In this paper, we present a novel diffeomorphic image registration network with efficient training process and inference. In contrast to traditional learning-based registration methods that are defined in the high dimensional imaging space, our model is fully developed in a low dimensional bandlimited space with the diffeomorphic property of transformation fields well preserved. Based on the fact that current networks are majorly designed in real-valued spaces, we are the first to develop a decoupled-network to solve the complex-valued optimization with the support of rigorous math foundations. Our model substantially lowers the computational complexity of the neural network and significantly reduces the time consumption on both training and testing, while preserving a comparable registration accuracy. The theoretical tools developed in our work is flexible/generic to a wide variety of the state-of-the-art networks, e.g., FCN, or RNN. To the best of our knowledge, we are the first to characterize the diffeomorphic deformations in Fourier space via network learning. This work also paves a way for further speeding up unsupervised learning for registration models.

{\small
\bibliographystyle{ieee_fullname}
\bibliography{egbib}
}

\end{document}